# The Empty Quadrant: AI Teammates for Embodied Field Learning

Hyein Kim[1] and Sung Park[1*]

[1] School of Data Science and Artificial Intelligence, Taejae University

*Corresponding author: Sung Park (sjp@taejae.ac.kr)

**Abstract.** For four decades, AIED research has rested on what we term the Sedentary Assumption: the unexamined design commitment to a stationary learner seated before a screen. Mobile learning and museum guides have moved learners into physical space, and context-aware systems have delivered location-triggered content -- yet these efforts predominantly cast AI in the role of information-delivery tool rather than epistemic partner. We map this gap through a 2×2 matrix (AI Role × Learning Environment) and identify an undertheorized intersection: the configuration in which AI serves as an epistemic teammate during unstructured, place-bound field inquiry and learning is assessed through trajectory rather than product. To fill it, we propose Field Atlas, a framework grounded in embodied, embedded, enactive, and extended (4E) cognition, active inference, and dual coding theory that shifts AIED's guiding metaphor from instruction to sensemaking. The architecture pairs volitional photography with immediate voice reflection, constrains AI to Socratic provocation rather than answer delivery, and applies Epistemic Trajectory Modeling (ETM) to represent field learning as a continuous trajectory through conjoined physical-epistemic space. We demonstrate the framework through a museum scenario and argue that the resulting trajectories -- bound to a specific body, place, and time -- constitute process-based evidence structurally resistant to AI fabrication, offering a new assessment paradigm and reorienting AIED toward embodied, dialogic human-AI sensemaking in the wild.

## 1   Introduction: Naming the Blind Spot

The AIED community has built an unparalleled tradition of marrying computational precision with pedagogical sensitivity [1,2]. From Carbonell's SCHOLAR (1970) through Cognitive Tutor to today's LLM-based dialogue agents [3,4], the field's greatest achievements share a common spatial arrangement: a stationary learner, a screen-mediated interface, and a system that delivers, diagnoses, and corrects knowledge. We term this pattern the Sedentary Assumption: a structural constraint on what AIED has been able to see, study, and support. The assumption manifests across system design (SCHOLAR through ChatGPT-based tutors all presuppose a fixed learner [5,6]), assessment (knowledge tracing and essay evaluation capture discrete products, not spatiotemporal processes [7,8]), and the AI-teammate concept itself (even the most collaborative AI visions remain anchored to shared digital workspaces [9]). Arroyo et al.'s [10] WearableLearning has begun to challenge this assumption from structured indoor settings, but unstructured field environments -- museums, archaeological sites, urban landscapes -- remain AIED's underexplored territory. AIED 2026's "From Tools to Teammates" theme [37] makes filling this quadrant timely.

This blind spot becomes urgent now because generative AI has destabilized product-based education: when LLMs produce polished essays in seconds, what a learner produces matters less than how they arrive at understanding -- the process of sensemaking itself [15,24]. The alternative, however, is not new. Over three decades of embodiment research -- from Varela et al. [11] through the embodied, embedded, enactive, and extended (4E) cognition framework [12] -- have established that cognition is constitutively shaped by bodily action, sensory experience, and coupling with physical environments. Hippocampal place cells bind memory to spatial context [13], and multisensory integration strengthens long-term encoding [14].

We make this blind spot visible through a 2×2 matrix (Fig. 1) that cross-classifies *AI Role* (Tool vs. Teammate) with *Learning Environment* (Screen-based vs. Field-based). Three quadrants are populated; the fourth -- *Field-Teammate* in unstructured field environments -- remains conceptually underarticulated. Mobile learning, museum guides, and location-based educational systems have moved learners into physical space [38], and context-aware recommenders have delivered place-triggered content. Yet these systems predominantly function as information-delivery tools: AI selects and presents content based on location, but does not engage the learner's own hypothesis formation or assess the sensemaking process itself. By 'empty quadrant' we mean not the absence of mobile educational technology, but the undertheorized intersection of three elements: (i) AI as an *epistemic teammate* constrained to Socratic prompting, (ii) *unstructured, place-bound field inquiry* in which the learning path is not predetermined, and (iii) *trajectory-based process assessment* that captures how understanding evolves. In other words, what is absent is not mobility, but the theorization of AI as a constrained epistemic provocateur whose primary function is to model and assess learning trajectories rather than deliver content. This paper advances Field Atlas as a theoretical articulation of this intersection, redefining the AI teammate as an *Epistemic Cartographer* -- an agent that captures, connects, and visualizes the learner's embodied sensemaking journey.

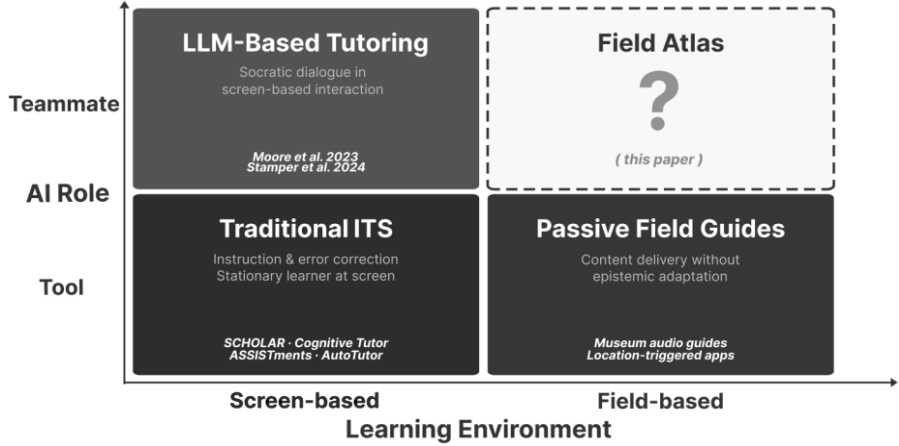

**Fig. 1.** The 2×2 AIED Landscape Matrix (AI Role × Learning Environment). Existing mobile and museum learning systems populate the Field-Tool quadrant. The undertheorized intersection -- AI as epistemic teammate in unstructured field environments -- is the territory Field Atlas proposes to fill.

## 2     From Instruction to Sensemaking

Table 1. Instruction paradigm vs. Sensemaking paradigm.

| Dimension | Instruction Paradigm (current dominant) | Sensemaking Paradigm (proposed) |
| --- | --- | --- |
| Learner's body | Sedentary | Ambulatory |
| Environment | Screen (structured) | Physical field (unstructured) |
| AI's role | Instructor / Corrector | Epistemic Cartographer |
| Unit of analysis | Discrete response | Epistemic trajectory |
| Assessment target | Product / Artifact | Process / Trajectory |

Table 1 crystallizes the paradigm shift. Its theoretical backbone is 4E cognition [11,12]: understanding is constitutively shaped by sensorimotor interaction with the physical world. In the field, embodied encounters generate prediction errors -- unexpected textures, spatial disorientation, perceptual surprise -- that drive deep belief updating [18,19], a dynamic formalized in active inference as epistemic foraging [16,17]. To preserve these encounters without disrupting them, Field Atlas harnesses dual coding [20,21]: simultaneous visual and verbal processing creates richer memory traces, but only when capture is *volitional* rather than mindless offloading [22,23]. Together, these principles motivate a shift from *Instruction* -- AI optimizing knowledge transmission -- to *Sensemaking* [15,24]: AI scaffolding the learner's own meaning-construction from unstructured experience.

These foundations connect to AIED Blue Sky discourse: Rismanchian & Doroudi [9] showed the field's conceptualization of AI-education interactions narrowed over time; Stamper et al. [3] demonstrated that ITS wisdom remains essential as LLMs transform the landscape; Borchers et al. [25] distinguished 'ethical AIED' from 'AIED ethics' -- a distinction we adopt in §4.

## 3  Field Atlas in Action

Rather than describing Field Atlas abstractly and then demonstrating it separately, we introduce each mechanism through a concrete scenario: Maya, a student, spends one hour in the American Wing (Gallery 760) of The Metropolitan Museum of Art. Fig. 2 provides the architectural overview; four mechanisms constitute the framework, unified by a single principle: learning in unstructured physical environments is best understood, supported, and assessed as a trajectory through epistemic space, authenticated by physical metadata.

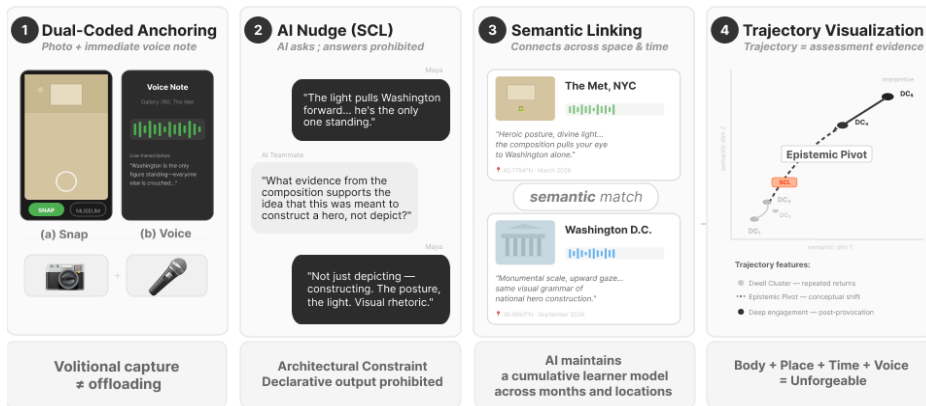

**Fig. 2.** Field Atlas four-layer architecture. Layers 1–2 operate in real time during field encounters; Layers 3–4 operate longitudinally, connecting experiences across space and time.

### 3.1  Dual-Coded Anchoring

The first mechanism preserves the cognitive richness of embodied encounters without disrupting them. Drawing on Barasch et al.'s [23] volitional-capture finding, Field Atlas specifies a minimal protocol: when the learner encounters an object of epistemic interest, they (a) perform intentional photography -- actively framing and zooming in -- and (b) immediately record a voice note articulating their observation or question. The photograph anchors attention; the voice note externalizes nascent reasoning; their temporal coupling harnesses dual coding [20] and hippocampal place-cell mechanisms that bind memory to spatial context [13] for robust memory formation.

Maya's first encounter is Emanuel Leutze's monumental Washington Crossing the Delaware (1851). She photographs the canvas from a low angle, zooming into Washington's upright stance amid huddled soldiers. Voice note: "Washington is the only figure standing -- everyone else is crouched or rowing. And the light hits him from the far shore, like it's pulling him forward. Is the whole composition built to make one person look inevitable?" This encounter becomes a dual-coded Data Card (photo + voice + GPS + timestamp), anchoring memory to a specific body, place, and moment.

### 3.2   AI as Epistemic Provocateur

The second mechanism specifies what the AI teammate does -- and critically, what it must not do. Under the Structured Cognitive Loop (SCL) [26], declarative answer generation is prohibited. The AI is constrained to generate epistemic provocations -- questions designed to perturb the learner's emerging interpretation. Returning to the painting, Maya records: "The flag is at the center but almost lost in the storm. And the ice chunks look more theatrical than real." The AI intervenes: "You described the light as 'pulling' Washington forward. What evidence from the physical composition supports the idea that this painting was meant to construct a hero, not merely depict a crossing?" This redirects attention from visual description to rhetorical interpretation, deepening the epistemic trajectory without prematurely terminating it.

### 3.3   Semantic Linking Across Space and Time

The third mechanism addresses what we term context collapse -- the failure to connect insights across discontinuous experiences [27]. As Data Cards accumulate, the system maintains a semantic network. When a new capture exhibits semantic similarity to prior Data Cards, the AI surfaces the connection as a provocation. Months after observing how composition and light construct heroism in Leutze's painting at the Met, Maya visits the Lincoln Memorial in Washington, D.C. She photographs the massive seated figure from below. The system intervenes: "You previously observed how a painting uses light and posture to make one person look inevitable. How does this architectural composition achieve a similar effect?" In doing so, AI becomes a longitudinal epistemic partner, enabling trajectories of sensemaking to extend across space and time.

### 3.4   Epistemic Trajectory Modeling (ETM)

The signature contribution is Epistemic Trajectory Modeling, adapting Tan's [30] ETM methodology -- building on Brohinsky et al.'s [29] trajectory visualization in Quantitative Ethnography [28] -- to embodied field learning. ETM provides a pipeline incorporating smoothing, dimensionality reduction, and trajectory-comparison metrics to model how understanding evolves as a continuous trajectory through conceptual space. The trajectory $T = \{(e_t, v_t) \mid t \in [0,T]\}$ captures conceptual development, with physical metadata (GPS, timestamp) serving as an authentication layer binding each point to a specific body, place, and moment.

Fig. 3 visualizes Maya's one-hour visit. Two trajectory features emerge: ***Epistemic Pivot***: following the AI provocation, her semantic embedding shifts sharply -- from descriptive vocabulary ("standing," "light," "ice") to interpretive vocabulary ("constructed heroism," "visual rhetoric," "propaganda"). This pivot illustrates how constraint-driven provocation reshapes the semantic landscape of inquiry. *Epistemic Velocity*: rapid conceptual movement in the first 15 minutes gives way to slower, deeper engagement in the final 30 minutes.

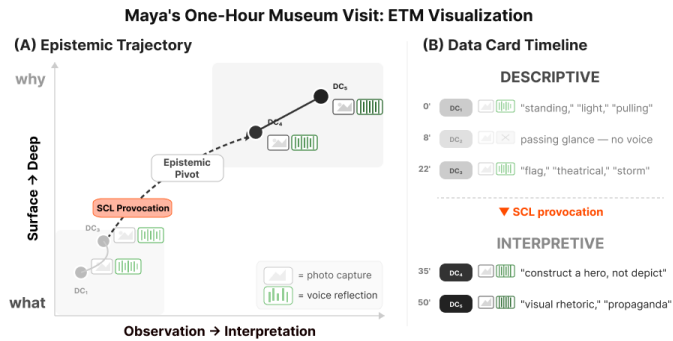

**Fig. 3.** Illustrative ETM visualization for Maya's museum visit, adapting Tan's [30] trajectory methodology. (A) Epistemic trajectory space: axes represent latent semantic dimensions derived from voice reflections; the pivot from descriptive to interpretive vocabulary following the SCL provocation is marked. Physical metadata anchors each point to a specific body, place, and time. (B) Data Card timeline showing the sequence of observations, the SCL provocation, and the shift toward interpretive framing.

The visualized trajectory operationalizes Hattie's [31] Visible Learning principle: impact is maximized when the learning process itself becomes visible to both teacher and learner. Critically, epistemic trajectories are structurally resistant to synthetic reconstruction. A trajectory is bound to a specific body moving through a specific environment at a specific time; fabricating it would require physically visiting the location, generating coherent visual inquiry patterns, and recording temporally aligned voice reflections in situ -- conditions that substantially raise the cost of fabrication [32]. The appropriate response to the AI ghostwriting crisis is therefore paradigmatic rather than technological (e.g., unreliable AI detection tools [33]).

The museum scenario demonstrates one configuration, but the framework generalizes to any physical environment (e.g., urban walking, nature trails, historical sites) and to any learner. Costa et al.'s [34] CLJ4AI extended AIED to citizen learning, and Field Atlas resonates because embodied sensemaking is not confined to classrooms. In this sense, Field Atlas repositions AIED from content optimization toward trajectory-aware learning design.

## 4    Open Challenges and Ethical Considerations

We identify four open challenges. Privacy: continuous GPS, photo, and voice recording raises serious concerns [35]; privacy-by-default principles and learner data ownership must be foundational design commitments. Equity: field-based learning presupposes mobility and device access, risking marginalization of constrained learners [36]; Field

Atlas is intended to expand AIED's scope rather than displace existing modalities. Metric validity: Epistemic Velocity and Epistemic Pivots are theoretically motivated but require rigorous empirical validation. Finally, a technologization paradox: we argue for the irreducibility of embodied experience while proposing digital capture. The map is not the territory, yet a well-designed map can enable wiser navigation [15].

## 5   Conclusion

This paper identifies the Sedentary Assumption as a structural blind spot in AIED and proposes Field Atlas to address the undertheorized intersection of epistemic AI partnership, unstructured field inquiry, and trajectory-based assessment. Our proposal is deliberately unconventional -- not a co-thinker in digital space, but an Epistemic Cartographer that captures, connects, and visualizes trajectories of understanding that emerge through embodied human experience. We suggest that a next frontier for AIED lies in developing a science of sensemaking grounded in embodied inquiry.

12. Newen, A., De Bruin, L., Gallagher, S. (eds.): The Oxford Handbook of 4E Cognition. Oxford Univ. Press, Oxford (2018)
13. O'Keefe, J., Nadel, L.: The Hippocampus as a Cognitive Map. Oxford Univ. Press, Oxford (1978)
14. Shams, L., Seitz, A.R.: Benefits of multisensory learning. Trends in Cognitive Sciences 12(11), 411–417 (2008)
15. Weick, K.E.: Sensemaking in Organizations. Sage Publications, Thousand Oaks (1995)
16. Friston, K.: The free-energy principle: A unified brain theory? Nature Reviews Neuroscience 11(2), 127–138 (2010)
17. Friston, K., Rigoli, F., Ognibene, D., Mathys, C., Fitzgerald, T., Pezzulo, G.: Active inference and epistemic value. Cognitive Neuroscience 6(4), 187–214 (2015)
18. Clark, A.: Surfing Uncertainty: Prediction, Action, and the Embodied Mind. Oxford Univ. Press, Oxford (2016)
19. Di Paolo, L.D., White, B., Guénin-Carlut, A., Constant, A., Clark, A.: Active inference goes to school: The importance of active learning in the age of large language models. Phil. Trans. R. Soc. B 379(1911), 20230148 (2024)
20. Paivio, A.: Imagery and Verbal Processes. Holt, Rinehart & Winston, New York (1971)
21. Mayer, R.E.: Multimedia Learning. 3rd edn. Cambridge Univ. Press, Cambridge (2020)
22. Henkel, L.A.: Point-and-shoot memories: The influence of taking photos on memory for a museum tour. Psychological Science 25(2), 396–402 (2014)
23. Barasch, A., Diehl, K., Silverman, J., Zauberman, G.: Photographic memory: The effects of volitional photo taking on memory for visual and auditory aspects of an experience. Psychological Science 28(8), 1056–1066 (2017)
24. Odden, T.O.B., Russ, R.S.: Defining sensemaking: Bringing clarity to a fragmented theoretical construct. Science Education 103(1), 187–205 (2019)
25. Borchers, C., Liu, X., Lee, H.H., Zhang, J.: Ethical AIED and AIED ethics: : Toward Synergy Between AIED Research and Ethical Frameworks. In: Olney, A.M., et al. (eds.) AIED 2024 Blue Sky, CCIS, vol. 2150, pp. 18–31. Springer, Cham (2024)
26. Kim, M.H.: Executable epistemology: The Structured Cognitive Loop as an architecture of intentional understanding. arXiv:2510.15952 (2025)
27. Godden, D.R., Baddeley, A.D.: Context-dependent memory in two natural environments: On land and underwater. British J. Psychology 66(3), 325–331 (1975)
28. Shaffer, D.W.: Quantitative Ethnography. Cathcart Press, Madison (2017)
29. Brohinsky, J., Marquart, C., Wang, J., Ruis, A.R., Shaffer, D.W.: Trajectories in epistemic network analysis. In: Ruis, A.R., Lee, S.B. (eds.) Advances in Quantitative Ethnography. ICQE 2020, CCIS, vol. 1312, pp. 106–121. Springer, Cham (2021)
30. Tan, Y.: Epistemic trajectory modeling: A method for modeling, visualizing, and comparing learning processes. In: Carmona, G., et al. (eds.) Seventh International Conference on Quantitative Ethnography: Conference Proceedings Supplement, pp. 19–21. ISQE, Mexico City (2025)
31. Hattie, J.: Visible Learning: A Synthesis of Over 800 Meta-Analyses Relating to Achievement. Routledge, London (2009)
32. Cotton, D.R.E., Cotton, P.A., Shipway, J.R.: Chatting and cheating: Ensuring academic integrity in the era of ChatGPT. Innovations in Education and Teaching Int. 61(2), 228–239 (2024)
33. Weber-Wulff, D., et al.: Testing of detection tools for AI-generated text. International Journal for Educational Integrity 19(1), 26 (2023)
34. Costa, L.F.C., et al.: CLJ4AI: Citizen Learning Journey for AI. In: Cristea, A.I., et al. (eds.) AIED 2025 Blue Sky, CCIS, vol. 2590, pp. 3–16. Springer, Cham (2025)